\documentclass[a4paper,11pt]{article}
\usepackage{pos}
\usepackage{wrapfig}
\usepackage[export]{adjustbox}
\usepackage{enumitem}

\title{Commissioning, Calibration, and Performance of the Cherenkov Telescope on EUSO-SPB2}
\ShortTitle{}

\author*[a]{Oscar Fernando Romero Matamala}
\onbehalf{for the JEM-EUSO Collaboration}

\affiliation[a]{School of Physics \& Center for Relativistic Astrophysics, Georgia Institute of Technology,\\
		     837 State Street NW, Atlanta, GA 30332-0430, USA}

\emailAdd{oromero@gatech.edu}

\abstract{The detection of astrophysical neutrinos by IceCube in the TeV-PeV energy range motivates the development of instruments for observing these particles at higher energies. Moreover, the detection of very-high-energy (VHE) neutrinos could potentially bring constraints on ultra-high energy cosmic rays (UHECRs) source models. Tau neutrinos skimming the Earth under a shallow angle can be detected through the decay of a tau resulting in an extensive air shower (EAS) in the atmosphere. The EAS can be detected by capturing some of the optical Cherenkov signal originating from the EAS particles. To assess the feasibility of the Earth-skimming technique from high altitudes, we developed a Cherenkov telescope which was deployed on the Extreme Universe Space Observatory Super Pressure Balloon 2 (EUSO-SPB2) from Wanaka, NZ, on May $13^{th}$. It is a precursor for the Probe of Extreme Multi-Messenger Astrophysics. The 1m diameter Cherenkov telescope for EUSO-SPB2 had a focal plane comprised of 512 silicon photomultipliers (SiPMs) covering a 6.4 x 12.8 square degree field of view coupled to a 100 MS/s readout based on the GET switch capacitor ring sampler. We discuss the calibration and commissioning of the telescope and its in-flight performance.}

\ConferenceLogo{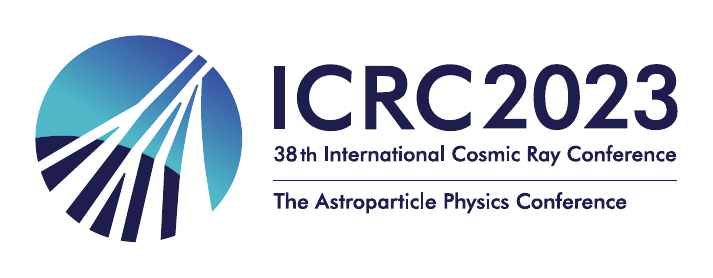}

\FullConference{%
The 38th International Cosmic Ray Conference (ICRC2023)\\
  26 July -- 3 August, 2023\\
  Nagoya, Japan}

\begin{document}
\maketitle

\section{Introduction}
The search for ultrahigh-energy (UHE; $>10^{7}$\,GeV) neutrinos is an exciting and timely topic in astroparticle physics. A lot of the interest can be traced to Icecube's detection of high-energy neutrinos \cite{Aarsten2013, ArstenMaxLike} and the possible identification of a distant black hole as a neutrino source \cite{Aartsen2014}. These measurements are strong motivations to extend measurements into the UHE regime. The interest in extending these measurements into higher energy can be seen in the increasing number of proposed experiments aiming to obtain detections in the UHE regime. Some of these current projects include Pierre-Auger, Ice Cube, ANITA, \cite{ThePierreAugerCollaboration2019LimitsObservatory, Cremonesi2019TheIce, Abbasi2011IceCubeSupernovae} and future ones like GRAND, ARIANNA, POEMMA, and Trinity. \cite{Anker2020, GRAND2018, Olinto2020, Brown2021Trinity:Neutrinos}.

\begin{figure}[!ht]
    \centering
    \begin{minipage}{.48\textwidth}
     \includegraphics[width=0.7\linewidth]{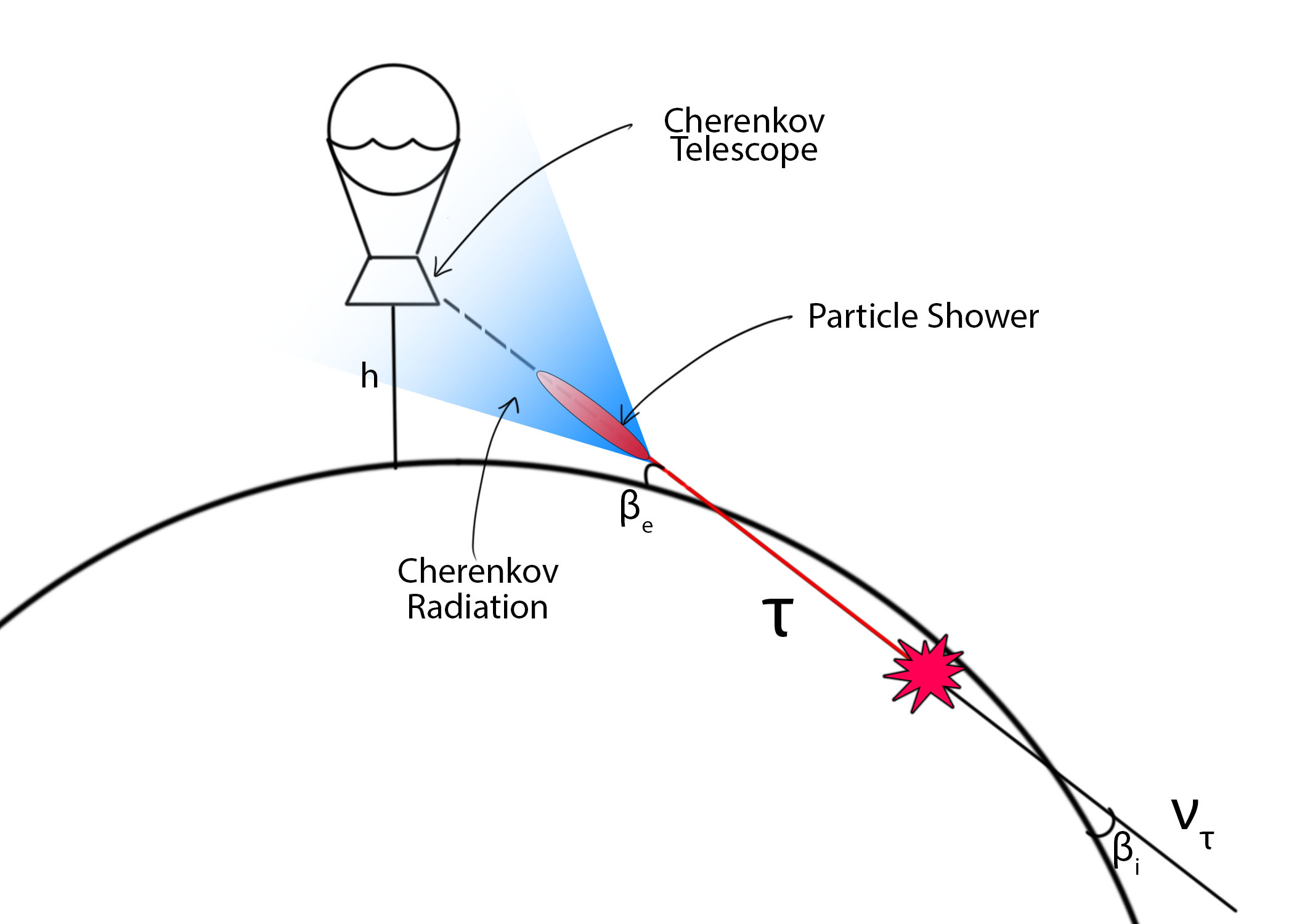}
     \caption{Illustrative sketch of the Earth-skimming technique for detecting tau neutrinos from EUSO-SPB2.}
g      \label{fig:earthskim}
    \end{minipage}\hfill
    \begin{minipage}{.48\textwidth}
        \centering
        \includegraphics[width=0.50\linewidth]{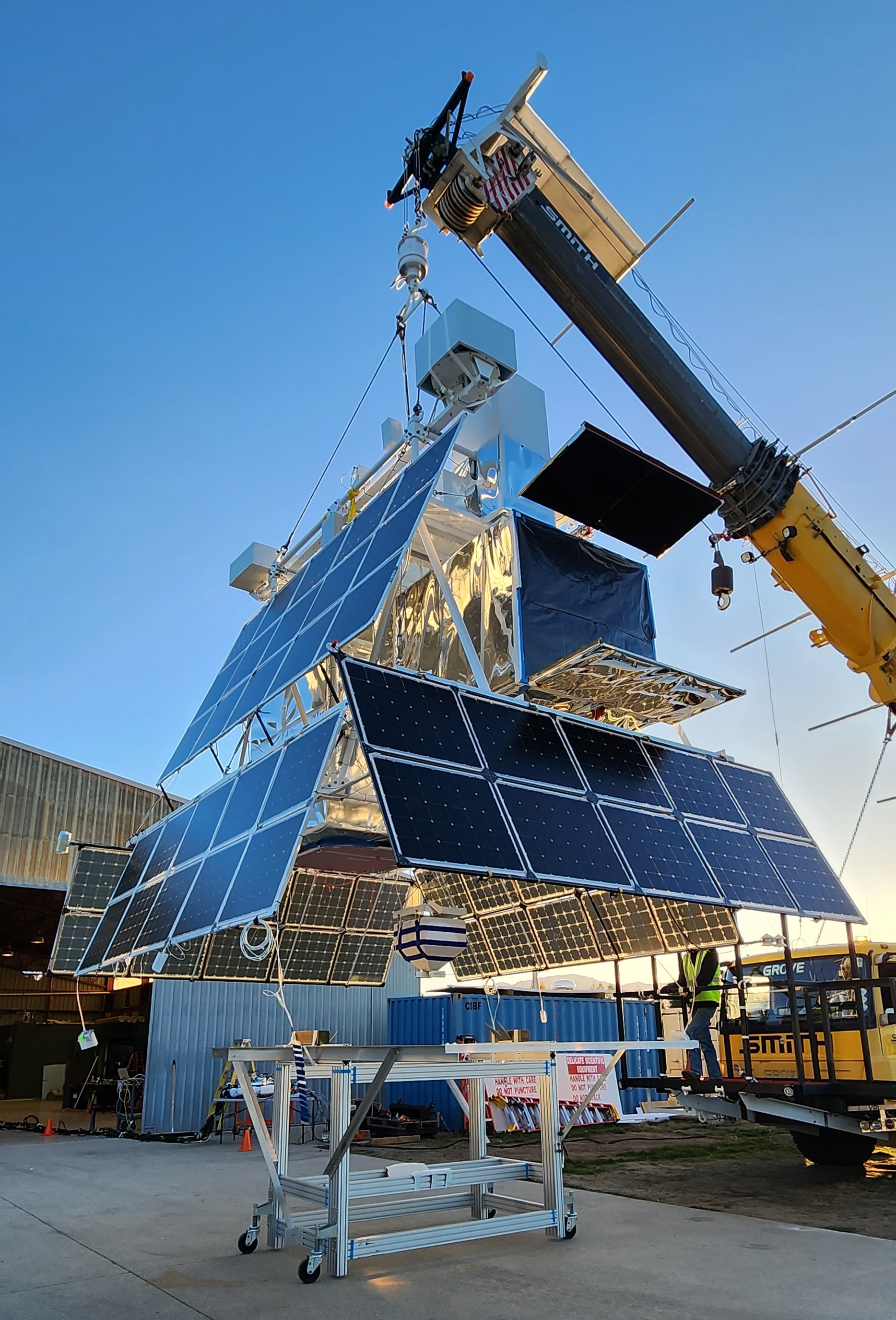}
        \caption{Picture from hang test before launch at Wanaka, NZ. The CT is at the front with its shutters open, and a cover to protect from sunlight while tests were done before launch.}
    \label{fig:telescope}
    \end{minipage}
\end{figure}

EUSO-SPB2 (Extreme Universe Observatory-Super Pressure Balloon 2) is an effort to extend observations into the UHE regime, by monitoring the Southern Hemisphere sky\cite{Adams2017}. The balloon payload consists of two air-shower imaging telescopes. One of them is a Cherenkov telescope \cite{Otte2019}, to observe the beamed optical Cherenkov radiation from billions of charged particles following the decay of a $\tau$-lepton in the atmosphere \cite{Reno2019} a schematic of the technique is shown in Figure \ref{fig:earthskim}. The second telescope is optimized to detect the fluorescence light from an air-shower following an Ultra High Energy Cosmic Ray (UHECR) interaction in the atmosphere \cite{Adams2017}. 

The CT The Cherenkov telescope (CT) uses Schmidt optics \cite{Adams2017} realized with segmented spherical mirror facets. The optical configuration of the mirrors is such that a bi-focal image is formed from parallel light. This is used to discriminate accidentals caused by the Night Sky Background (NSB). A photograph of the fully integrated payload is shown in Figure \ref{fig:telescope}. The camera on board is comprised of 512 Silicon Photo-Multipliers (SiPM) packaged in matrices each with 16 SiPMs (4x4 arrangement). The SiPMs are then connected to a Multiple Use SiPM Integrated Circuit (MUSIC), which is a readout ASIC specifically designed for CTA \cite{Gomez2016MUSIC}. The MUSIC amplifies the signal while also providing a discriminator. On trigger, the ASIC for General Electronics TPC (AGET) and the ASIC Support \& Analog Digital conversion (ASAD) \cite{Anvar2011AGETExperiments} digitize the incoming signal. The AGET amplifies, detects and stores the signal so that the ASAD can perform the digitization. The AGET uses 512 time bins each with $10\ ns$ time width. The ASAD has a 12-bit ADC which reads out the stored analog signal in the AGET and digitizes it for the Concentration Board (CoBo) to compress, time stamp, and package for storage in the computer.  In preparation for flight, the telescope was tested extensively to obtain relative and absolute calibrations for its different components in the signal chain.
\section{Calibration}
\subsection{Photon Detection Efficiency}
A central component to the telescope's performance is the SiPMs present in the camera. Characterizing their response and calibrating it is central to understanding observations done with the CT. One of the most important characteristics of the SiPMs is determining the photon-detection efficiency (PDE) of the SiPMs \cite{Otte2016CharacterizationPhotomultipliers}. It was determined that the performance of each SiPM within a matrix did not vary considerably. Therefore, just one pixel was used for measuring the characteristics of each of the matrices. 
A setup as described in \cite{Otte2016CharacterizationPhotomultipliers} for measuring the PDE of a SiPM was used. The PDE of the detector is recorded at different bias voltages and the voltage to obtain 90\% breakdown probability is chosen. This bias is the nominal operational voltage and is the one at which all calibrations and measurements are taken. 


\subsection{Spectral Response}
\begin{figure}
    \centering
    \begin{minipage}{.40\textwidth}
     \includegraphics[width=0.9\textwidth]{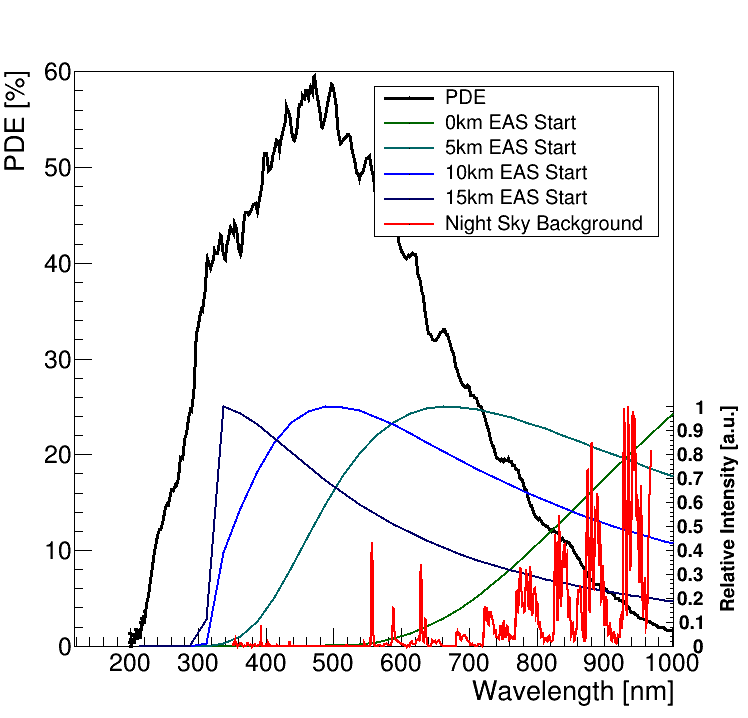}
    \caption{PDE of Hamamatsu S14521 compared to a $1^\circ$ Earth emergence shower and the measured NSB. Both spectra have been amplitude normalized. Oscillations in the PDE are from diffraction patterns in the Silicon substrate of the SiPM.}
    \label{fig:Spectral_Response}
    \end{minipage}\hfill
    \begin{minipage}{.58\textwidth}
        \centering
        \includegraphics[width=\textwidth]{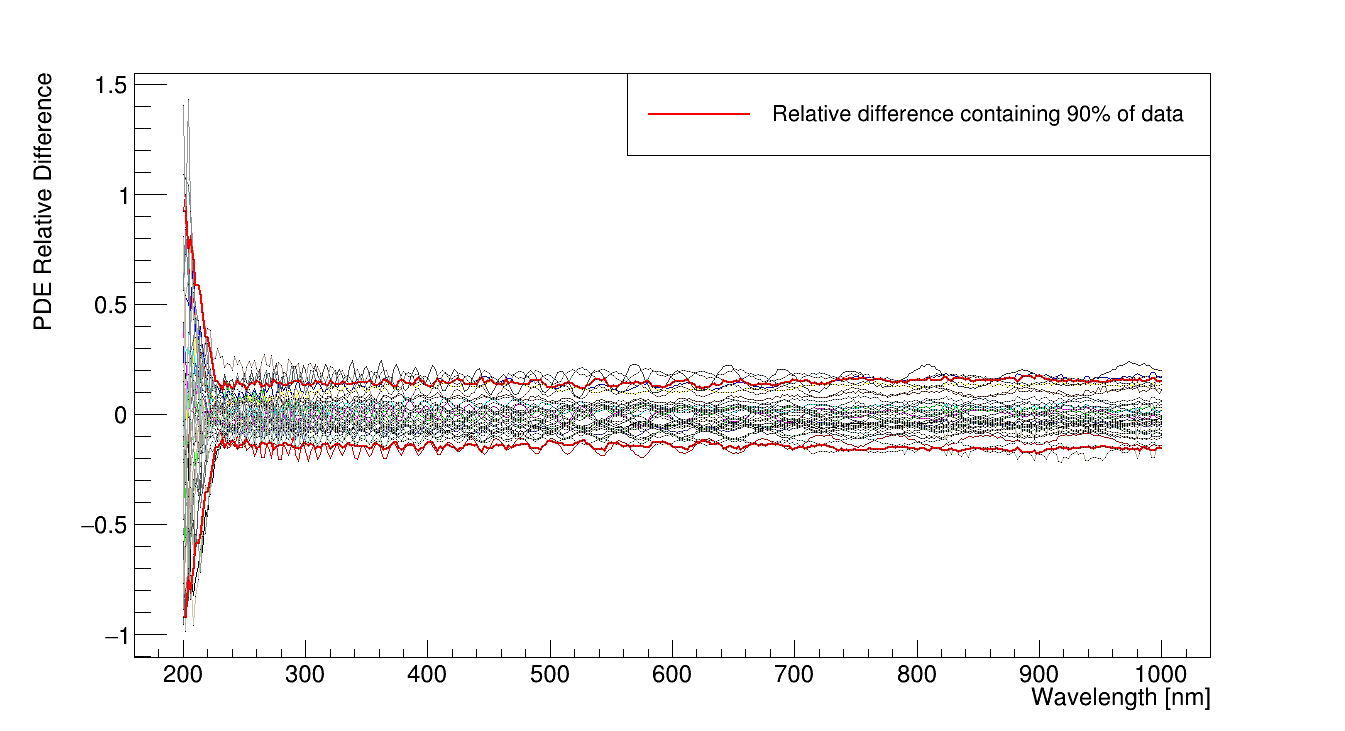}
        \caption{Dispersion of the relative difference between each characteristic spectral response. The distribution of the spectral responses has a standard deviation of around 18\%}
    \label{fig:PDE_Dist}
    \end{minipage}
    
\end{figure}

The sensitivity of these detectors to different wavelengths of light had to be characterized to further compare their performance. The spectral response setup uses a monochromator and a white light Xe lamp. The light beam is then concentrated by a lens and split into 2 beams. One of them goes to a calibrated monitor diode and the other to the SiPM \cite{Otte2016CharacterizationPhotomultipliers}. The spectrum ranging from 200 nm to 1000 nm is scanned and a spectral response of the detector is obtained. The spectral response of a single SiPM is shown in Figure \ref{fig:Spectral_Response} and the matrices' dispersion is shown in Figure \ref{fig:PDE_Dist}.

\subsection{System Response Linearity}
\begin{figure}
    \centering
    \begin{minipage}{.48\textwidth}
     \includegraphics[width=0.90\textwidth]{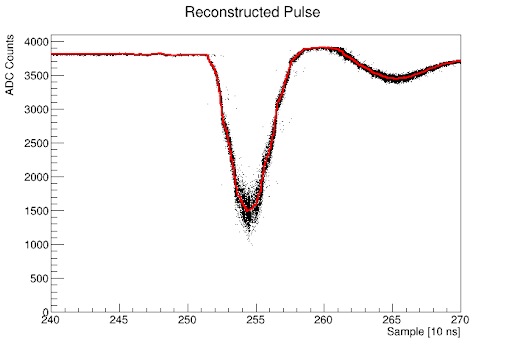}
    \caption{Oversampled digitized pulse. Cherenkov photons are emitted in times shorter than 10 ns, which would create a signal similar to what is shown. }
    \label{fig:DigiSignal}
    \end{minipage}\hfill
    \begin{minipage}{.48\textwidth}
        \centering
        \includegraphics[width=0.95\textwidth]{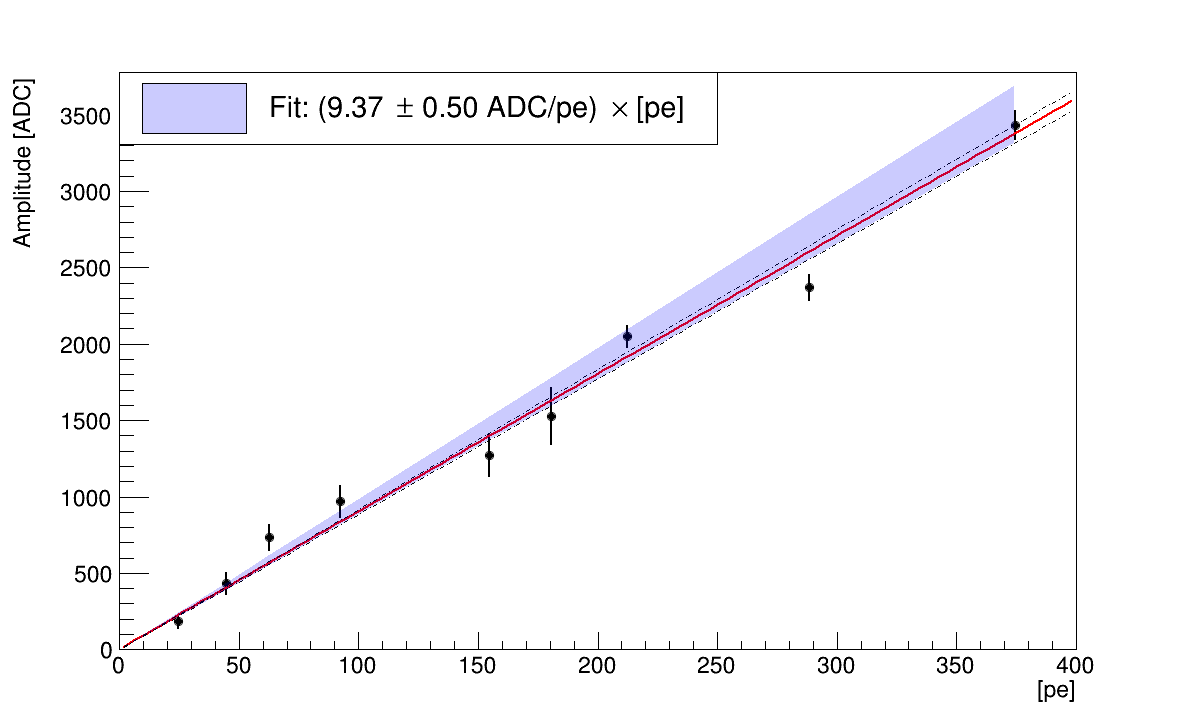}
        \caption{Linearity of a single pixel and dispersion of the whole system. The fitted line and points plotted correspond to a single pixel in the camera. The shaded blue region corresponds to the average calibration of the whole camera and its standard deviation.}
    \label{fig:lin}
    \end{minipage}
\end{figure}

Using a picosecond laser and the full readout system, the amplitude in ADC counts for different laser intensities were recorded. Using an oscilloscope the single photo-electron amplitude was recorded along with the amplitude in mV for each light intensity. Thus, a direct relation could be drawn from pe to mV and then from mV to ADC. Since the flashed laser is a picosecond laser, it is safe to assume that all the photons are bunched up and arrive at the same time at the SiPM. A sample of a digitized signal is shown in Figure \ref{fig:DigiSignal}. This setup would allow us to calibrate the observed ADC amplitudes and directly relate them to the number of photo-electrons. The result of this measurement is shown in Figure \ref{fig:lin}.

From this result, it is possible to infer the number of photo-electrons in a signal from the amplitude of the digitized signal.

\subsection{Discriminator Linearity}
\begin{wrapfigure}{r}{0.45\textwidth}
    \centering
    \includegraphics[width=0.45\textwidth]{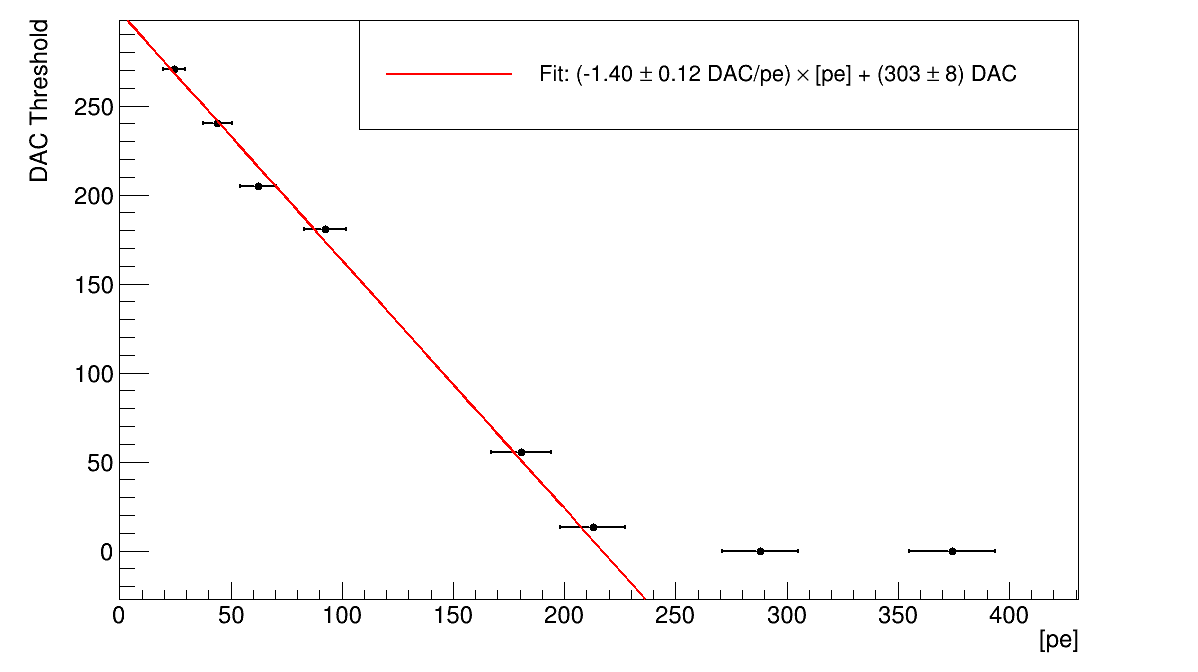}
    \caption{Linearity profile of the MUSIC discriminator. After about 200 pe the discriminator saturates and the trigger threshold is always exceeded.}
    \label{fig:DiscLin}
\end{wrapfigure}
The discriminator used for triggering was the one packaged with the MUSICs \cite{Gomez2016MUSIC}. To be able to relate this to the number of photo-electrons in the SiPM, the linearity of the discriminator had to be characterized. The same setup as for determining the linearity of the digitizing system was used. The trigger threshold of the discriminators was changed monotonically from a high threshold to a low one. At each LED amplitude a trigger efficiency curve was obtained. It is expected that at 50\% efficiency the threshold is set at the middle of the Poisson oscillations. Using this 50\% efficiency point the trigger threshold vs number of photo-electrons can be determined for the camera as shown in Figure \ref{fig:DiscLin}.

\section{Commissioning}
\subsection{Flat Fielding}
An important consideration in characterizing the linearity of both the system and the discriminator is that the camera must be flat-fielded. As observed in Figure \ref{fig:PDE_Dist} the PDE of each matrix varies considerably. Along with the PDE, there is also a dispersion in the gain of each SiPM. These 2 factors influence the observed digitized amplitude. To obtain a camera that responds homogenously to the same light source, a flat-fielding procedure was carried out.
To achieve this, all the SiPMs in the camera were biased to the same value, this value had been identified as the average at which the breakdown probability equals 90\%. Once all SiPMs have the same bias, the camera was pointed to a screen that acts as a Lambertian surface, and a light source was flashed to the screen. A relative difference with respect to the average was calculated using the digitized amplitude. This was then related to a fraction of the over-voltage, as the gain is directly proportional to it \cite{Otte2016CharacterizationPhotomultipliers}. Small corrections in the bias voltage for each SiPM were applied by modifying the voltage offset in the MUSIC readout channels. This fine-tuning has the effect of causing a hardware flat-field of the camera, such that the product of the PDE and the gain is constant across the camera (as both are affected by the over-voltage).

\begin{figure}
    \centering
    \begin{minipage}{.48\textwidth}
    \includegraphics[width=\textwidth]{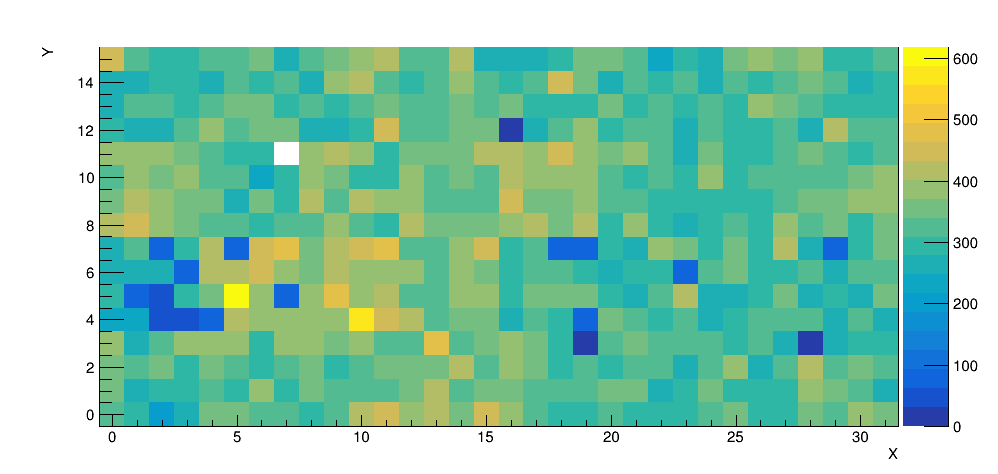}
    \caption{Average of 100 flashed events. The dark pixels had bad connections with the digitizer and the connectors were fixed before flight.}
    \label{fig:FlatFieldCam}
    \end{minipage}\hfill
    \begin{minipage}{.48\textwidth}
        \centering
        \includegraphics[width=\textwidth]{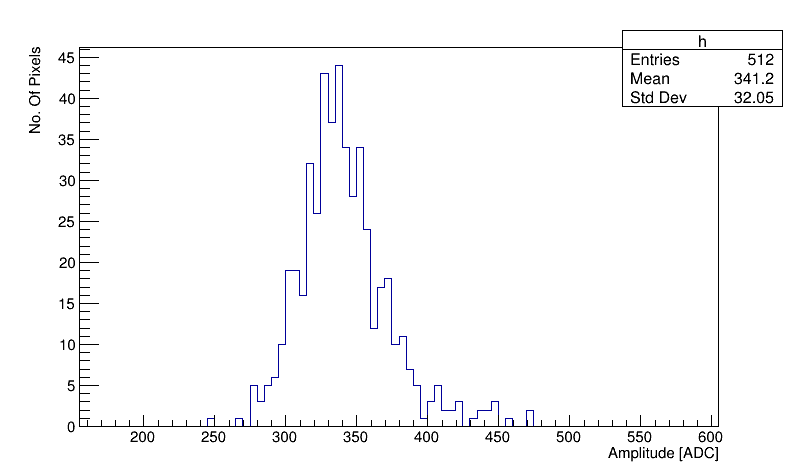}
        \caption{Dispersion of the mean of each of the SiPMs. }
    \label{fig:FlatFieldDisp}
    \end{minipage}
\end{figure}

Once the camera has been flat-fielded using this procedure, it is possible to assume that the camera will respond equally to the same light intensity. The dispersion of the non-flat-fielded camera response was 18\% which was improved to a rough 10\% relative error of the average across the whole flat-fielded camera. The response of the flat-fielded camera to the Lambertian surface used for flat-fielding is shown in Figure \ref{fig:FlatFieldCam}. The distribution of the average ADC amplitude of the digitized signal for ~100 flashes is shown in Figure \ref{fig:FlatFieldDisp}. If the distribution is Poisson limited, then the standard deviation should be smaller than what is observed. This widening in the distribution is due to electrical noise, the dispersion within each matrix, and the limited bias voltage adjustment of each SiPM as the effective dynamic range of the MUSIC only allows for 600 mV adjustment.

\begin{figure}[h]
    \centering
    \begin{minipage}{.48\textwidth}
    \includegraphics[width=\textwidth]{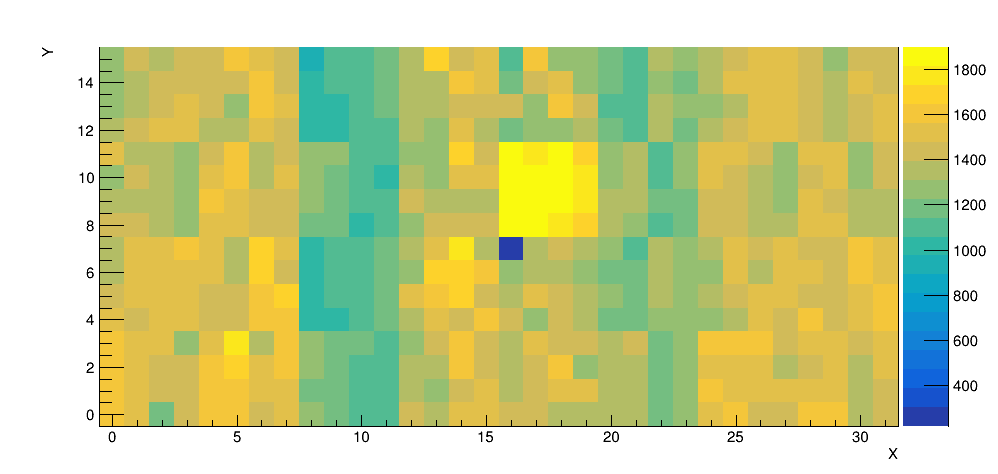}
    \caption{Camera image of the HLED flashes after amplitude extraction. The 2 cold matrix columns are readily seen along with a bright spot probably also caused by a geometric effect from the mirrors.}
    \label{fig:HLEDCam}
    \end{minipage}\hfill
    \begin{minipage}{.48\textwidth}
        \centering
        \includegraphics[width=\textwidth]{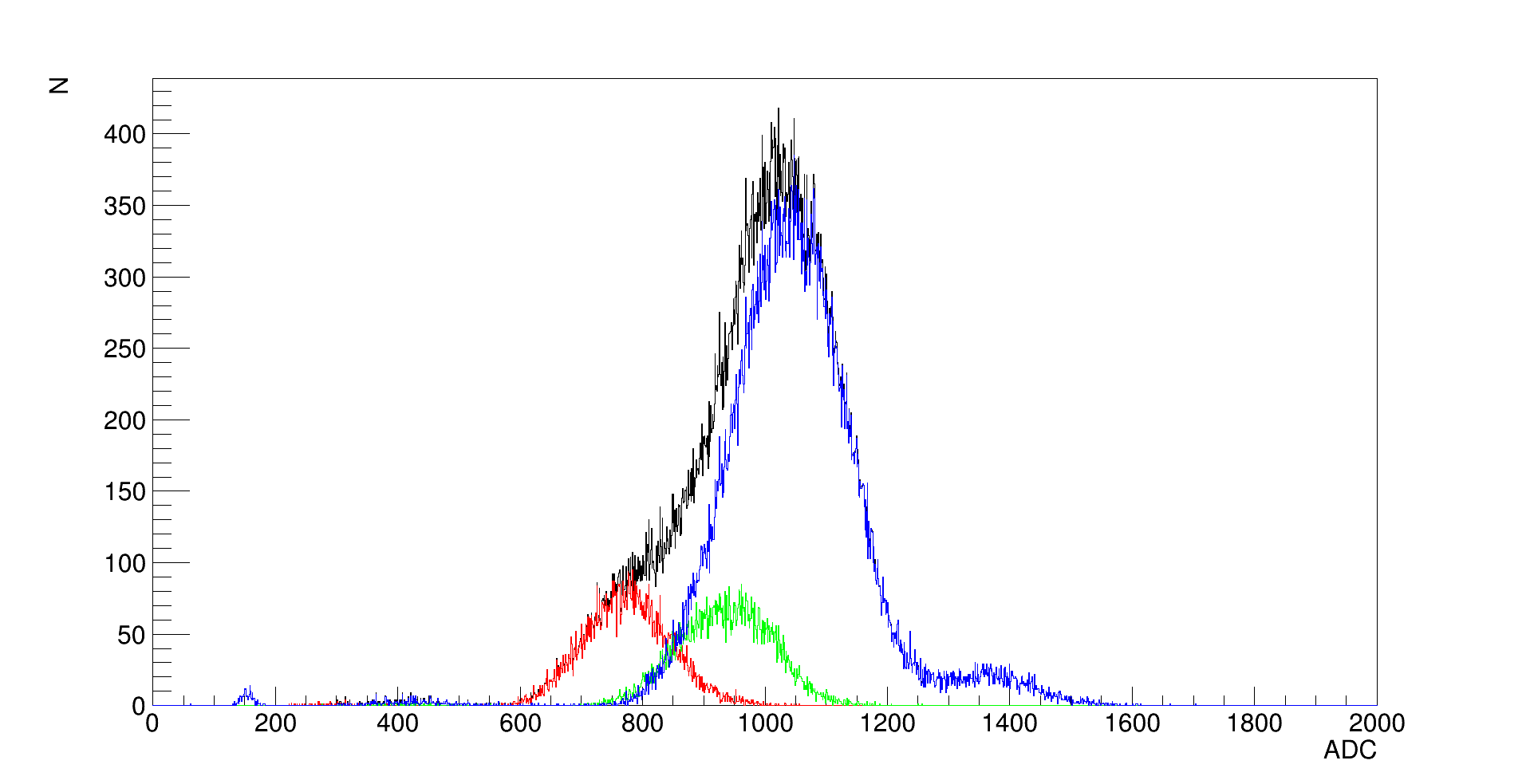}
        \caption{Distribution of the HLED amplitude after 1000 flashes. The blue distribution is the unobscured regions of the camera. The red and green correspond to the obscured regions. The black distribution is the sum of all the regions.}
        \label{fig:HLED_Dist}
    \end{minipage}
\end{figure}

The CT was flown with two Health LEDs (HLED), which allowed evaluating the state of the SiPMs in the camera. The setup of the HLED was such that the dispersed light of the HLED was shown from close to the focal plane, to the mirrors. It was identified that the gaps between the mirror segments were also projected to the camera, causing 2 distinct columns in the camera to be obscured from one of the HLEDs. One of the HLED flashes captured is shown in Figure \ref{fig:HLEDCam}, 
A similar test as the flat-fielding procedure was done by flashing the HLEDs in the telescope, while the aperture was closed (effectively making it a dark box). Once the mirror obscuration was taken into account, 3 distinct distributions were observed in the camera. By flashing the HLED 1000 times and extracting the amplitude, it was observed that the system was Poisson limited. This result is shown in Figure \ref{fig:HLED_Dist}, where the 3 distinct regions (the 2 columns that only see one HLED and the rest of the camera) are separated.

With this, the performance of the camera had been characterized and the instrument was flight-ready. The camera could be operated within its linear range and its performance was understood.

\subsection{Trigger Scans}
During flight, multiple trigger scans were performed to characterize the background trigger rate. The CT was the first to attempt observation of Cherenkov light from a suborbital vantage point. As such, information about the background light and expected trigger rates by accidentals and actual events was very limited. The discriminator threshold was lowered until the noise floor was reached. This allowed us to understand the background light observed when looking above and below the horizon. At the time of this writing, the data obtained from these scans are still being analyzed to determine actual rates for both accidentals and candidate events.
\section{Performance}
The performance of the CT is being evaluated as of this writing. The CT had 2 nights of observation aboard SPB2. About 30,000 candidate events were recorded during this time. These are not the totality of the triggered events observed in flight. The full data set was pre-processed to discard all events that did not meet the bifocal condition. Since the discriminators do a logical OR of 8 pixels at a time, it is possible to create a trigger without a true bifocal signal. Therefore, every event was checked for compliance with the bifocal optics (the signal is contained in 2 pixels separated by one pixel). These candidate events can still pass both tests without being actual cosmic-rays. The analysis and calibration of these events is still ongoing.

\section{Conclusion}
For the first time, a Cherenkov Telescope was flown at 33km altitude and captured around 30,000 candidate events. The pre-flight calibration and commissioning of the instrument allows us to confidently evaluate the obtained data from this short but fruitful flight. The in-depth study of the electronics comprising the CT camera, will aid in the analysis of the data and prove that the technique used is viable and can produce meaningful scientific data even in extreme conditions. The detailed analysis of the electronics will serve future missions like POEMMA in implementing and understanding their Cherenkov Telescopes, opening the avenue to novel contributions to the UHE and VHE cosmic ray field.

\section{Acknowledgements} 
\noindent

The authors acknowledge the support by NASA awards 11-APRA-0058, 16-APROBES16-0023, 17-APRA17-0066, NNX17AJ82G, NNX13AH54G, 80NSSC18K0246, 80NSSC18K0473, 80NSSC19K0626, 80NSSC18K0464, 80NSSC22K1488, 80NSSC19K0627 and 80NSSC22K0426, the French space agency CNES, National Science Centre in Poland grant n. 2017/27/B/ST9/02162, and by ASI-INFN agreement n. 2021-8-HH.0 and its amendments. This research used resources of the US National Energy Research Scientific Computing Center (NERSC), the DOE Science User Facility operated under Contract No. DE-AC02-05CH11231. We acknowledge the NASA BPO and CSBF staffs for their extensive support. We also acknowledge the invaluable contributions of the administrative and technical staffs at our home institutions.
\bibliographystyle{JHEP}
\bibliography{references,Thesis}

\clearpage
\newpage
{\Large\bf Full Authors list: The JEM-EUSO Collaboration\\}

\begin{sloppypar}
{\small \noindent
S.~Abe$^{ff}$, 
J.H.~Adams Jr.$^{ld}$, 
D.~Allard$^{cb}$,
P.~Alldredge$^{ld}$,
R.~Aloisio$^{ep}$,
L.~Anchordoqui$^{le}$,
A.~Anzalone$^{ed,eh}$, 
E.~Arnone$^{ek,el}$,
M.~Bagheri$^{lh}$,
B.~Baret$^{cb}$
% ...
}
\end{sloppypar}
\vspace*{.3cm}

{ \footnotesize
\noindent
% ...
$^{cb}$ Universit\'e de Paris, CNRS, AstroParticule et Cosmologie, F-75013 Paris, France\\
% ...
$^{ed}$ Istituto Nazionale di Fisica Nucleare - Sezione di Catania, Italy\\
% ...
$^{eh}$ INAF - Istituto di Astrofisica Spaziale e Fisica Cosmica di Palermo, Italy\\
% ...
$^{ek}$ Istituto Nazionale di Fisica Nucleare - Sezione di Torino, Italy\\
$^{el}$ Dipartimento di Fisica, Universit\`a di Torino, Italy\\
% ...
$^{ep}$ Gran Sasso Science Institute, L'Aquila, Italy\\
% ...
$^{ff}$ Nihon University Chiyoda, Tokyo, Japan\\ 
% ...
$^{ld}$ University of Alabama in Huntsville, Huntsville, AL, USA\\
$^{le}$ Lehman College, City University of New York (CUNY), NY, USA\\
% ...
$^{lh}$ Georgia Institute of Technology, USA\\
% ...
}

\end{document}